\documentclass[onecolumn,amsmath,amssymb,showpacs]{revtex4}

\usepackage{graphicx}
\usepackage{dcolumn}
\usepackage{bm}
\usepackage{epsf}
\usepackage{amsmath}
\usepackage{amssymb}
\usepackage{color}

\newcommand{\JPCM}{{\it J. Phys.: Condens. Matter\/} }    

\newcommand{\SST}{{\it Semicond. Sci. Technol.} }
\newcommand{\SUST}{{\it Supercond. Sci. Technol.} }

\newcommand{\NC}{{\it Nuovo Cimento\/} }

\newcommand{\PL}{{\it Phys. Lett.} }
\newcommand{\PR}{{\it Phys. Rev.} }
\newcommand{\PRL}{{\it Phys. Rev. Lett.} }

\newcommand{\SSC}{{\it Solid State Commun.} }

\begin{document}

\title[Effect of interband interactions of phonon and charge fluctuation]{Effect of interband interactions of phonon and charge fluctuation on superconducting parameters of MgB$_2$}

\author{O C Abah\footnote{\textit{Present address}: Department of Nanoscience, Delft University of Technology, Lorentzweg 1, 2628 CJ Delft, The Netherlands.}, G C Asomba and C M I Okoye}

\address{Department of Physics and Astronomy, University of Nigeria, Nsukka, Nigeria.}
\begin{abstract}
We have investigated the effect of unconventional pairing mechanism in MgB$_2$  using a two-band model within the framework of Bogoliubov-Valatin formalism. The approach incorporates the intraband s-wave interaction in the s- and p-bands, as well as interband s-wave interaction between them.  The analysis assumes the pairing interaction matrix 
comprises of attractive  electron-phonon, charge fluctuation and repulsive electron-electron (Coulomb) interactions to account for superconductivity in MgB$_2$. The model is
used to estimate the transition temperature and isotope effect exponent as well as to elucidate the importance of interband contributions in the superconductivity of the system.\\
\textit{Published version:} \SUST 23, 045031 (2010).

\end{abstract}

\maketitle

\section{Introduction}

\begin{sloppypar} 
The discovery of superconductivity in magnesium diboride, MgB$_2$, at a transition temperature, $T_c \sim$ 39 K \cite{Nagamatsu} has attracted a lot of studies both theoretically and experimentally in the field of condensed matter physics. This intermetallic material was associated to conventional superconductors based on pairing by electron-phonon coupling but its remarkable high-$T_c$ \cite{Vinod2007} cannot be explained using the conventional (Bardeen-Copper-Schrieffer (BCS)) theory \cite{BCS}. Also, the experimentally \cite{Budko2001,Hinks2001} observed shift in the isotope exponent reveals the importance of phonon contribution to the $T_c$.

Electronic bands structure calculation \cite{Belashchenko2001,Kortus2001} predicted the existence of 2D covalent in-plane ($\sigma$-band) and 3D metallic type inter-layer ($\pi$-band) conducting bands at the Fermi level, E$_F$ for this system. The $\sigma$-band is strongly coupled with phonons within the honeycombed boron layer by electrons while electrons in the $\pi$-band can support only weak coupling.

The multi-band gap nature of superconductivity in MgB$_2$ was first predicted theoretically by Liu et al. \cite{Liu2001} and have been observed in various experiments; such as, tunneling spectroscopy \cite{Iavarone2002,Giubileo2001}, point-contact spectroscopy \cite{Szabo2001}, specific heat measurements \cite{Bouquet1} and magneto-Raman spectroscopy \cite{Blumberg2007}. The two-band nature of superconductivity in MgB$_2$ is already well established \cite{Xi2008} and the BCS mechanism have been supported by photoemission spectroscopy \cite{Takahashi2001} and scanning tunneling microscopy \cite{Karapetrov2001} among others. Also, the specific heat analysis performed by W\"alte et al. \cite{Walte2006} have revealed the weak interband coupling scenario in MgB$_2$ and they show that the data are best explain within the two-band BCS theory of superconductivity. 

Choi et al. \cite{Choi2002} used the Eliashberg formalism to show that the $\sigma$-bonding states possess an average energy gap of 6.8 meV while the $\pi$-states have weak pairs with an average energy gap of 1.8 meV. Pickett \cite{Pickett2002} explained the double gap as having two transition temperatures, one at 45 K ($\Delta_{\sigma}$) and the other at 15 K ($\Delta_{\pi}$), which compromise to results in a transition temperature (for the bulk material) of 39 K \cite{Nagamatsu}. Putti et al. \cite{Putti2008}, Nicol and Carbotte \cite{Nicol2005} provided a good illustration on how the two-bands integrate to results into a single $T_c$ due to strong interband processes.  That is, the two-gaps merge to one isotropic BCS gap \cite{Mazin2003} and possess one critical temperature as a result of strong interband scattering \cite{Liu2001,Putti2008,Nicol2005,Ferrando2007}.  
However, the merging of the two bands does not result in the superconducting properties of a one-band superconductor because of the presence of strong anisotropy coupling as well as both intra- and inter-band scattering  that affect the superconducting properties  \cite{Nicol2005,Putti2008}.

In MgB$_2$, the boron isotope exponent, $\beta$(B) is only significant while the Mg isotope exponent, $\beta$(Mg) is small but non-zero. Bud\'ko et al. \cite{Budko2001} measured $\beta$(B) $= 0.26$ and the same value was recently obtained by Brotto et al. \cite{Brotto2008}. Hinks et al. \cite{Hinks2001} measured $\beta$(B) to be 0.30 and $\beta$(Mg) to be 0.02, resulting in a total isotope exponent, $\beta$, equals 0.32 for MgB$_2$. \"Ord et al. \cite{Ord2002} calculated $\beta = 0.34$ using electron-phonon and Coulomb interactions in the $\sigma$-bands as well as the interband scattering of intraband pairs within a two-band model. Choi et al. \cite{Choi2003} have calculated $\beta$(B) $= 0.32$ and $\beta$(Mg) $= 0.03$ using isotropic Eliashberg theory. Also, Calandra \cite{Calandra2007} group found that inclusion of strong electron-phonon coupling and Coulomb repulsion effect in one-band or two-bands Migdal-Eliashberg approach suppress the isotope exponent of boron, $\beta$(B), to a value of about 0.4 - 0.45. Studies have suggested that the low isotope effect is mainly due to phonon anharmonicity \cite{Choi2002PRB} and interband Coulomb repulsion \cite{Ord2002,Abah2009}. 

Despite the fact that the isotope effect experiment supports the phonon mediated BCS type superconductor, the observed small value of $\beta$ cannot be explained by the conventional BCS theory \cite{BCS}. Also, large amount of effort devoted to the study of this phenomenon have not provided a clear understanding of the reduced isotope effect \cite{Calandra2007} and the pairing mechanism   \cite{Vinod2007}. Among the nonconventional models that have been proposed for explaining the superconductivity in MgB$_2$ are electronic mechanism mediated by collective excitations pairing \cite{Zhukov2001,Sharapov2002}, bipolarons  \cite{Alexandrov2001}, and the electronic resonance-valence-bond (RVB) model  \cite{Baskaran2002}.

Ku et al. \cite{Ku2002} showed that the presence of collective excitations, consisting of coherent charge fluctuation between Mg and B sheets affects the optical properties of the material. Shortly after discovery, electron energy-loss spectroscopy (EELS) \cite{Keast2001,Yu2001} suggested that a special electronic contribution may affect the superconductivity of MgB$_2$. First principle calculation predicted that the excitations arise from interband transitions in MgB$_2$ \cite{Zhukov2001}. Combined EELS and ab initio calculations of the plasmon structure in MgB$_2$ have shown a peak at 2.4 eV and several peaks for higher energies \cite{Keast2005}. Optical experiment \cite{Guritanu2006} revealed that the intense peak is due to the transition from the $\sigma$- to the $\pi$-band. The optical study went further to suggest that this collective electronic modes affect the color of MgB$_2$. That is, polarization of light is bluish silver for E$\|$ab and yellow for E$\|$c. Another optical experiment suggested that the multi-band picture is necessary to understand the optical spectra of MgB$_2$ \cite{Kakeshita2006}. First principles calculations of the excitations spectra in MgB$_2$ involving all the three symmetry momentum directions have confirmed the existence of the long-lived collective excitations and its presence in the region of optical frequencies strongly affects the optical properties \cite{Balassis2008}. Recently, Silkin et al. \cite{Silkin2009} demonstrated that the unknown long-lived collective mode corresponds to coherent charge fluctuations between the boron $\sigma$ and $\pi$ band ($\sigma \pi$ mode) as well as having a periodic sine-like dispersion for energies below 0.5 eV. Varshney and Nagar \cite{Varshney2007} have employed a model involving collective charge fluctuation, within the Eliashberg formalism, to calculate some superconducting parameters of MgB$_2$ without the inclusion of interband interactions. 

The present study is motivated by the predictions of collective coherent charge fluctuation from first principles calculations \cite{Zhukov2001,Ku2002,Balassis2008,Silkin2009}, EELS \cite{Keast2001,Yu2001} and optical \cite{Guritanu2006,Kakeshita2006} experiments as well as the fact that MgB$_2$ is conclusively agreed to be a two-band BCS type superconductor with exceptional high $T_c$ (see reviews, Ref. \cite{Vinod2007,Xi2008}). Therefore, we employ the two-band BCS model within the Bogoliubov-Valatin formalism and naively assume that the pairing interaction matrix comprises attractive electron-phonon, repulsive Coulomb and attractive electron-plasmon interactions to study the superconducting parameters of MgB$_2$. This is to elucidate the role of charge fluctuation on the system and the interband contribution to the superconducting properties of MgB$_2$. This kind of pairing mechanism have been used by Tewari and Gumber \cite{Tewari} within one-band BCS model to study the effect of plasmons on the yttrium and lanthanum based superconductors.
The paper is organized as follows; in section 2, we present the model. Then, section 3 will be devoted to results and analysis. Finally, section 4 is the conclusion.
\end{sloppypar}

\section{Model}
\label{model}

\begin{sloppypar}
In accord with the original fomulation by Suhl et al. \cite{Suhl1959}, on two-band superconductivity, and more recently, studies on two-band model of MgB$_2$ superconductor \cite{Abah2009,Kristoffel2003,Udoms2005}, the effective Hamiltonian of the system can be written as;
\end{sloppypar}

\begin{eqnarray}
H  =   \sum_{iks} \epsilon_{ik}c^{\dagger}_{iks}c_{iks}        +\sum_{ikk'}V_{ikk'}c^{\dagger}_{ik'\uparrow}c^{\dagger}_{i-k'\downarrow}c_{i-k\downarrow}c_{ik\uparrow} \nonumber   \\
       +  \sum_{kk'}V_{\pi\sigma kk'} \left(  c^{\dagger}_{\pi k \uparrow}c^{\dagger}_{\pi -k \downarrow}c_{\sigma -k' \downarrow}c_{\sigma k' \uparrow}  
 +  c^{\dagger}_{\sigma k \uparrow}c^{\dagger}_{\sigma -k \downarrow}c_{\pi -k' \downarrow}c_{\pi k' \uparrow} \right)
 \label{Hamiltonian}
\end{eqnarray}
where $\epsilon_{ik}$ are kinetic energies of the two $\left(i=\pi,\sigma\right)$bands measured relative to the Fermi level, $k(k')$ is the Bloch wave vector, $V_{ikk'}$ are the intraband potential matrices, s is spin index, $\uparrow$ or $\downarrow$, $c^{\dagger}_{iks}$ ( $c_{iks}$) are the creation (annihilation) operators, for $ i^{th}$ band, and $V_{\pi\sigma k k'}$ is the interband interaction. Employing the standard Bogoliubov-Valatin transformation \cite{Bogoliubov,Valatin} in equation (\ref{Hamiltonian}), the linearized gap equations (T$_c$ equation) can be written as
\begin{equation}
\Delta_{\pi k} = - \sum_{k'} V_{\pi kk'} \frac{\Delta_{\pi k'}}{2\epsilon_{\pi k}} \left(1-2f\left(\epsilon_{\pi k'}\right) \right) - \sum_{k'} V_{\pi \sigma kk'} \frac{\Delta_{\sigma k}}{2\epsilon_{\sigma k'}}\left(1-2f\left(\epsilon_{\sigma k'}\right) \right) 
\label{gap1}
\end{equation}
\begin{equation}
\Delta_{\sigma k} = - \sum_{k'} V_{\sigma kk'} \frac{\Delta_{\sigma k'}}{2\epsilon_{\sigma k}} \left(1-2f\left(\epsilon_{\sigma k'}\right) \right) - \sum_{k'} V_{\pi \sigma kk'} \frac{\Delta_{\pi k}}{2\epsilon_{\pi k'}}\left(1-2f\left(\epsilon_{\pi k'}\right) \right) 
\label{gap2}
\end{equation}             
where $\Delta_{ik}$ are the gap parameters for two bands, $\pi$ and $\sigma$, $f\left(\epsilon_{ik}\right)$ are the Fermi-Dirac occupation number for quasiparticle state to energies $\epsilon_{ik}$ above the Fermi level. 

In the present model, we shall assume the pairing interaction matrix $(V_{ikk'})$ are made up of contributions from the attractive electron-phonon (V$_{ph}$), repulsive electron-electron(Coulombic) (V$_{c}$) and electron-plasmon interaction (V$_{pl}$). Therefore;
\begin{equation}
V_{ikk'}=\left\{ \begin{array}{lll}
                    -V_{iph}, &  if & \mbox{$ \omega \le \omega_{ph}$} \\
                     V_{ic},        &  if & \mbox{$\omega_{ph} \le \omega \le \omega_{c}$} \\ 
                    -V_{ipl},             & if  &  \mbox{$\omega_{c} \le \omega \le \omega_{pl}$} \\
                    0					& if & \omega > \omega_{pl}
                    \end{array}
           \right.      
\label{interaction-matrix}
\end{equation}
where $\omega_{ph}$ and $\omega_{pl}$ are the cut-off frquencies for the  electron-phonon and electron-plasmon (collective excitation), respectively; and $\omega_{c}$ is that for the electron-electron interaction which corresponds to the Fermi energy of the system. Also, $\omega_{pl} > \omega_{c} > \omega_{ph}$ and $i=(\pi, \sigma$ and $\pi\sigma \, (\sigma\pi))$. 

\begin{sloppypar}
Employing equation (\ref{interaction-matrix}) in equations (\ref{gap1}) and (\ref{gap2}) and replacing the summation over $k'$ by integration over energy, $\epsilon_{ik'}$ (using $k_B = \hbar = 1$), we obtain
\end{sloppypar}
\begin{eqnarray}
\Delta_{\pi kj} = -N(0)\int\left( -V_{\pi ph}+V_{\pi c}-V_{\pi pl}\right) \frac{\Delta_{\pi kj}}{2\epsilon_{\pi k'}} \left( 1 - 2f\left( \epsilon_{\pi k'}\right) \right) d \epsilon_{\pi k'} \nonumber \\
-N(0)\int\left( -V_{\pi\sigma ph}+V_{\pi\sigma c}-V_{\pi\sigma pl}\right) \frac{\Delta_{\sigma kj}}{2\epsilon_{\sigma k'}} \left( 1 - 2f\left( \epsilon_{\sigma k'}\right) \right) d \epsilon_{\sigma k'}
\label{gap-pi}
\end{eqnarray}
\begin{eqnarray}
\Delta_{\sigma kj} = -N(0)\int\left( -V_{\sigma ph}+V_{\sigma c}-V_{\sigma pl}\right) \frac{\Delta_{\sigma kj}}{2\epsilon_{\sigma k'}}\left( 1 - 2f\left( \epsilon_{\sigma k'}\right) \right) d \epsilon_{\sigma k'} \nonumber \\
-N(0)\int\left( -V_{\sigma\pi ph}+V_{\pi c}-V_{\sigma\pi pl}\right) \frac{\Delta_{\pi kj}}{2\epsilon_{\pi k'}}\left( 1 - 2f\left( \epsilon_{\pi k'}\right) \right) d \epsilon_{\pi k'}
\label{gap-sigma}
\end{eqnarray}
where j = ph, c, and pl.\\
We solve the resulting equations following standard procedure \cite{Su1990,Okoye1998} which involves separating the phonon and non-phonon parts. From equation (\ref{gap-pi}) for the $\pi$-band, we obtained three equations each representing the electron-phonon, Coulombic and electron-plasmon parts as follows: The electron-phonon part;
\begin{eqnarray}
\Delta_{\pi kph} = N(0)V_{\pi ph}\int_{-\omega_{ph}}^{\omega_{ph}} \Delta_{\pi kph} \frac{( 1 - 2f\left( \epsilon_{\pi k'}\right)}{{2\epsilon_{\pi k}}} d \epsilon_{\pi k'} \, \nonumber\\ + N(0)V_{\pi \sigma ph}\int_{-\omega_{ph}}^{\omega_{ph}} \Delta_{\sigma kph}\frac{(1 - 2f(\epsilon_{\sigma k'}))}{2\epsilon_{\sigma k'}}.
\label{pi-1}
\end{eqnarray}
Integrating and re-arranging equation (\ref{pi-1}) gives
\begin{equation}
(1 - \lambda_{\pi} Z_1) \Delta_{\pi ph} - \lambda_{\pi\sigma} Z_1\Delta_{\sigma ph} - \Delta_{\pi c} = 0
\end{equation}
where $Z_{1} = \ln\left(\frac{1.14\omega_{ph}}{T_c}\right), \lambda_{\pi} = N(0)V_{\pi ph},  \lambda_{\pi \sigma} =N(0)V_{\pi \sigma ph}$ and $\Delta_{\pi c}$ is integration constant.

The electronic part is given by
\begin{eqnarray}
\Delta_{\pi kc} = -N(0)V_{\pi c}\left(\int_{-\omega_c}^{-\omega_{ph}} \Delta_{\pi kc} + \int_{-\omega_{ph}}^{\omega_{ph}} \Delta_{\pi kph} + \int_{\omega_{ph}}^{\omega_{c}} \Delta_{\pi kc}\right) \frac{(1 - 2f(\epsilon_{\pi k'}))}{2\epsilon_{\pi k'}} d\epsilon_{\pi k'} \nonumber\\
-N(0)V_{\pi\sigma c}\left(\int_{-\omega_c}^{-\omega_{ph}} \Delta_{\sigma kc} + \int_{-\omega_{ph}}^{\omega_{ph}} \Delta_{\sigma kph} + \int_{\omega_{ph}}^{\omega_{c}} \Delta_{\sigma kc}\right) \frac{(1 - 2f(\epsilon_{\sigma k'}))}{2\epsilon_{\sigma k'}} d\epsilon_{\sigma k'}.
\end{eqnarray}
This yields,
\begin{equation}
(1 + \mu_{\pi}Z_2)\Delta_{\pi c} + \mu_{\pi}Z_1\Delta_{\pi ph} + \mu_{\pi \sigma}Z_1\Delta_{\sigma ph} + \mu_{\pi\sigma}Z_2\Delta_{\sigma c} - \Delta_{\pi pl} = 0
\end{equation}
where $Z_2 = \ln\left(\frac{\omega_c}{\omega_{ph}}\right), \, \mu_{\pi} = N(0)V_{\pi c}$ and $\mu_{\pi\sigma} = N(0)V_{\pi\sigma}$.

Finally, the electron-plasmon part gives
\begin{eqnarray}
\Delta_{\pi kpl} = N(0)V_{\pi pl}\left( \int_{-\omega_{pl}}^{-\omega_c} \Delta_{\pi kpl} + \int_{-\omega_c}^{-\omega_{ph}} \Delta_{\pi kc} + \int_{-\omega_{ph}}^{\omega_{ph}} \Delta_{\pi kph}  
+ \int_{\omega_{ph}}^{\omega_{c}} \Delta_{\pi kc} + \int_{\omega_c}^{\omega_{pl}} \Delta_{\pi pl}\right) \frac{(1 - 2f(\epsilon_{\pi k'}))}{2\epsilon_{\pi k'}} d\epsilon_{\sigma k'} \nonumber \\
 - N(0)V_{\pi\sigma pl}\left( \int_{-\omega_{pl}}^{-\omega_c} \Delta_{\sigma {kpl}} + \int_{-\omega_c}^{-\omega_{ph}} \Delta_{\sigma kc} + \int_{-\omega_{ph}}^{\omega_{ph}} \Delta_{\sigma kph} 
+ \int_{\omega_{ph}}^{\omega_{c}} \Delta_{\sigma kc} + \int_{\omega_c}^{\omega_{pl}} \Delta_{\sigma kpl}\right) \frac{(1 - 2f(\epsilon_{\sigma k'}))}{2\epsilon_{\sigma k'}} d\epsilon_{\sigma k'}.
\end{eqnarray}

Then, we obtain
\begin{equation}
(1 - \lambda_{\pi pl}Z_3)\Delta_{\pi pl} - \lambda_{\pi pl}Z_1\Delta_{\pi ph} -\lambda_{\pi pl}Z_2\Delta_{\pi c} - \lambda_{\pi\sigma ph} Z_1\Delta_{\sigma ph} - \lambda_{\pi\sigma pl}Z_2\Delta_{\sigma c} - \lambda_{\pi\sigma pl}Z_3\Delta_{\sigma pl} = 0
\end{equation}
where $Z_{3} = \ln\left( \frac{\omega_{pl}}{\omega_{c}}\right), \, \lambda_{\pi pl} = N(0)V_{\pi pl}$ and  $\lambda_{\pi\sigma pl} = N(0)V_{\pi \sigma pl}$.

Similarly, for the $\sigma$-band, (equation \ref{gap-sigma}), we obtain another set of three homogeneous equations corresponding to $\Delta_{\sigma ph}, \, \Delta_{\sigma c}$ and $\Delta_{\sigma pl}$ respectively. Thus, we can write the resulting six homomogenous equations in matrix form as;
\begin{equation}
\left( 
\begin{array}{cccccc}
 -1 & 1-\lambda_{\pi}Z_{1} & 0 & 0 & -\lambda_{\pi \sigma}Z_{1} & 0\\
1+\mu_{\pi}Z_{2} & \mu_{\pi}Z_{1} & -1 & \mu_{\pi \sigma}Z_{2}  & \mu_{\pi \sigma}Z_{1} & 0\\
-\lambda_{\pi pl}Z_{2} & -\lambda_{\pi pl}Z_{1} & 1-\lambda_{\pi pl}Z_{3} & -\lambda_{\pi \sigma pl}Z_{2} & -\lambda_{\pi \sigma pl}Z_{1} & -\lambda_{\pi \sigma pl}Z_{3}\\
0 & -\lambda_{\pi \sigma}Z_{1} & 0 & -1 & 1- \lambda_{\sigma}Z_{1} & 0\\
\mu_{\pi \sigma}Z_{2} & \mu_{\pi \sigma}Z_{1} & 0 & 1 +\mu_{\sigma}Z_{2} & \mu_{ \sigma}Z_{1} & -1\\
-\lambda_{\pi \sigma pl}Z_{2} & -\lambda_{\pi \sigma pl}Z_{1} & -\lambda_{\pi \sigma pl}Z_{3} & -\lambda_{\sigma pl}Z_{2} &  -\lambda_{\sigma pl}Z_{1} & 1 - \lambda_{\sigma pl}Z_{3}
\label{6x6matrix}
\end{array}
\right)
\left(
\begin{array}{c}
\Delta_{\pi kc} \\
\Delta_{\pi kph} \\
\Delta_{\pi kpl} \\
\Delta_{\sigma kc}\\
\Delta_{\sigma kph}\\
\Delta_{\sigma kpl}
\end{array}
\right) =0
\end{equation}
where,
 $\lambda_{\sigma} = N(0)V_{\sigma ph}, \hspace{.5cm} \mu_{\sigma} = N(0)V_{\pi c}$ and $\hspace{.5cm} \lambda_{\sigma} = N(0)V_{\sigma pl}$.

For non-trivial solutions, the determinant of the 6 $\times$ 6 matrix must vanish. To proceed, following the theory of multiband superconductivity \cite{Suhl1959}, we consider the case when both $\sigma$- and $\pi$- bands vanish simultaneously at superconducting transition temperature. For MgB$_2$, this vanishing of the gaps is due to interband phonon coupling between the $\sigma$- and $\pi$- bands \cite{Putti2008,Nicol2005,Golubov2002}. Therefore, equation (\ref{6x6matrix}) reduces to;
 
\begin{equation}
\left( 
\begin{array}{ccc}
 -1 & 1-(\lambda + \lambda_{\pi \sigma})Z_{1} & 0 \\
1 +(\mu + \mu_{\pi \sigma})Z_{2} & (\mu + \mu_{\pi \sigma})Z_{1} & -1\\
-(\lambda_{pl} +\lambda_{\pi \sigma pl})Z_{2} & -(\lambda_{pl} + \lambda_{\pi \sigma pl})Z_{1} & 1 - (\lambda_{pl} +\lambda_{\pi \sigma pl})Z_{3} 
\end{array}
\right)
\left(
\begin{array}{c}
\Delta_{0} \\
\Delta_{\alpha} \\
\Delta_{\theta}
\end{array}
\right) =0
\end{equation}
where
\begin{eqnarray}
 \Delta_{\pi kc} = \Delta_{\sigma kc} \equiv \Delta_{0}; \hspace{0.5cm} \Delta_{\pi kph} = \Delta_{\sigma kph} \equiv \Delta_{\alpha}; \hspace{0.5cm} \Delta_{\pi kpl} = \Delta_{\sigma kpl} \equiv \Delta_{\theta}; \nonumber \\
\lambda_{\pi} = \lambda_{\sigma} \equiv \lambda;\hspace{1.5cm}\lambda_{\pi pl} = \lambda_{\sigma pl} \equiv \lambda_{pl};\hspace{1.5cm} \mu_{\pi} = \mu_{\sigma} \equiv \mu.
\end{eqnarray} 
Solving the determinant of the 3 $\times $ 3 matrix, we obtain an expression for the transition temperature as; 
\begin{equation}
 T_{c} = 1.14 \omega_{ph} \exp \{-\frac{1}{\lambda_{eff}}\}
 \label{temperature}
\end{equation} 
where the effective coupling parameter,
\begin{equation}
 \lambda_{eff}=(\lambda +\lambda_{\pi \sigma}) - \frac{1}{Z_{2}+\frac{1}{(\mu + \mu_{\pi \sigma}) -\lambda^{\ast}_{\pi \sigma pl}}}
\end{equation}
and the renormalized electron-plasmon contribution parameter,
\begin{equation}
\lambda^{\ast}_{\pi \sigma pl} = \frac{\lambda_{pl} + \lambda_{\pi \sigma pl}}{1 - (\lambda_{pl} + \lambda_{\pi \sigma pl})Z_{3}}.
\end{equation}  
Equation (\ref{temperature}) is the expression for the transition temperature. The well known McMillan \cite{McMillan1968} expresion in two-band and two-square-well can be recovered from equation (\ref{temperature}) if we ignore the contribution of electron-plasmon ($\lambda_{pl} = 0 = \lambda_{\pi \sigma pl}$). Also, the one-band BCS model result is recovered if we neglected the electron-plasmon contributions as well as all interband contributions.

\begin{sloppypar}
We proceed to test the suitability of the model in account for the phonon contribution to pairing by exploring the isotope effect. The isotope exponent ($\beta$) can be derived from the expression for
$T_c$ given in equation (\ref{temperature}). The isotope effect exponent, $\beta$ is given by 
\begin{equation}
\beta = - \frac{d\ln T_{c}}{d\ln M}.
\end{equation} 
where $T_{c}$ is given by equation (\ref{temperature}) and M is the ionic mass.
 
Taking T$_{c}$ $\propto$ M$^{-\beta}$ and recalling that $\omega_{ph}$ $\propto$ M$^{-\frac{1}{2}}$ while $\omega_{pl}$ $\propto$ M$^0$, we obtain 
\begin{equation}
\beta = \frac{1}{2}\left\lbrace 1-\left\lbrace \frac{1}{\lambda_{eff}\left\lbrace Z_{2}+\frac{1}{(\mu + \mu_{\pi \sigma}) -\lambda^{\ast}_{\pi \sigma pl}}\right\rbrace }\right\rbrace^{2} \right\rbrace  
\label{isotope}
\end{equation} 
Equation (\ref{isotope}) is the expression for the isotope exponent ($\beta$) obtained using a two-band BCS gap equations and assuming that the electron-phonon, repulsive Coulomb and electron-plasmon (charge fluctuation) interacting mechanism are simultaneously present in the system. It can be easily seen that the presence of interband contributions influences the value of $\beta$.
\end{sloppypar}

\section{Results and Discussion}
\label{results}
\begin{sloppypar}
To understand the effect of the assumed pairing mechanism in MgB$_2$, we proceed to estimate the transition temperature and isotope effect exponent using the available data in the literature. At this point, we have to point out that the repulsive Coulomb coupling parameter is associated to the renormalized Coulomb pseudopotential, $\mu^{\ast} = \frac{\mu}{1+\mu Z_2}$ which depends on the scaling factor Z$_2 = \ln(\omega_c/\omega_{ph})$. Although, this can be seen as a crude approximation but we are encouraged by the common believe that the effects of Coulomb screening are drastically reduced by retardation effects due to the different energy scales for electrons and phonons. The effect of Coulombic repulsion is an important concept which is mainly fixed as an adjustable parameter in first principle calculations. A good account of this have been described in Ref. \cite{Varshney2007}.
\end{sloppypar}

\begin{sloppypar}
In the present analysis, the electron-phonon and the electron-plasmon cut-off frequency are $\omega_{ph} = 812$ K \cite{Goncharov2001} and $\omega_{pl} = 2.2$ eV \cite{Zhukov2001} respectively, while the frequency of Coulomb repulsive parameter, $\omega_c = 750$ meV \cite{Nicol2005}. Assuming an arbitrary total intraband coupling (comprising of electron-phonon and electron-plasmon coupling) $\lambda + \lambda_{pl} = 0.45$, we estimated the transition temperature, based on equation (\ref{temperature}) as 38.80 K for intra- (inter-)band Coulomb pseudopotential, $\mu^{\ast} (\mu^{\ast}_{\pi\sigma}) = 0.18\, (0.07)$ and interband parameters of $\lambda_{\pi\sigma pl} = 0.09$ and $\lambda_{\pi\sigma} = 0.08$.

\end{sloppypar}
\begin{sloppypar}
\begin{figure}[h]
{\scalebox{1.0}
{\includegraphics[width=10cm]{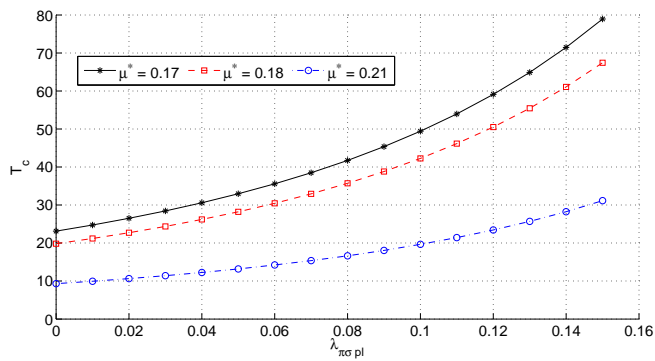}}}
\caption{Variation of  transition temperature ($T_c$) with interband electron-plasmon  interaction ($\lambda_{\pi \sigma pl}$) for different values of $\mu^{\ast}$}
\label{tempVsPlasmon}
\end{figure}

\begin{figure}[!]
{\scalebox{1.0}
{\includegraphics[width=10cm]{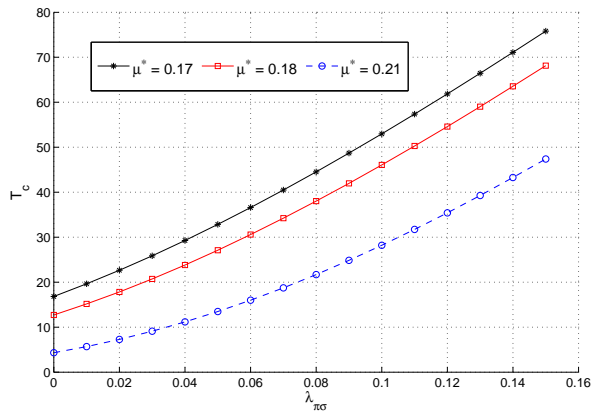}}}
\caption{Variation of  transition temperature ($T_c$) with interband electron-phonon  interaction ($\lambda_{\pi \sigma}$) for different values of $\mu^{\ast}$}
\label{tempVsPhonon}
\end{figure}

Figure (\ref{tempVsPlasmon}) shows the variation of $T_c$ with $\lambda_{\pi\sigma pl}$ for the above set of parameters and three different values of  $\mu^{\ast}$, that is 0.17, 0.18, and 0.21. This assumed $\mu^{\ast}$ compares well with the Golubov et al. \cite{Golubov2002} values of $\mu^{\ast}_{\pi \pi} = 0.17$ and $\mu^{\ast}_{\sigma\sigma} = 0.21$. It can be observed that $T_c$ increases with $\lambda_{\pi\sigma pl}$ but decreases with $\mu^{\ast}$. Also, $T_c \simeq 38.80$ K when $\lambda_{\pi\sigma pl} = 0.09$ for $\mu^{\ast} = 0.18$, $\mu^{\ast}_{\pi\sigma} = 0.07$ and $\lambda_{\pi\sigma} = 0.08$. And the effective interband coupling, $\lambda_{\pi\sigma} + \lambda_{\pi\sigma pl} = 0.08 + 0.09 = 0.17$.
 Also in figure (\ref{tempVsPlasmon}), we can see that $\lambda_{\pi\sigma pl} = 0$ corresponds to $T_c \sim 20$ K (for $\mu^{\ast} = 0.18$), which clearly shows the important of interband interaction in the high transition temperature of the MgB$_2$.

The variation of $T_c$ with the interband electron-phonon interaction is shown in figure \ref{tempVsPhonon}. The lower values of $\lambda_{\pi\sigma}$ yield unphysical value of $T_c$ irrespective of $\mu^{\ast}$. The effect of electron-phonon coupling strength on the renormalized repulsive Coulomb parameter, $\mu^{\ast}$ is consistent with the conventional superconductor. This shows that contribution of interband interactions of phonon and plasmon is crucial in high-$T_c$ of MgB$_2$.
\end{sloppypar}

\begin{sloppypar}
We proceed to analyze the isotope effect of the superconducting MgB$_2$ based on equation (\ref{isotope}). Employing the same parameters used to estimate the transition temperature, $T_c \simeq 38.80$ K, we obtain an exponent, $\beta \simeq 0.43$. This is in good aggrement with the theoretical calculated values by Choi et al. \cite{Choi2003} and Calandra et al. \cite{Calandra2007} but not close to the experimental measured values by Bud'ko et al. \cite{Budko2001} and Hinks et al.  \cite{Hinks2001}. 
\begin{figure}[h]
{\scalebox{1.0}
{\includegraphics[width=9cm]{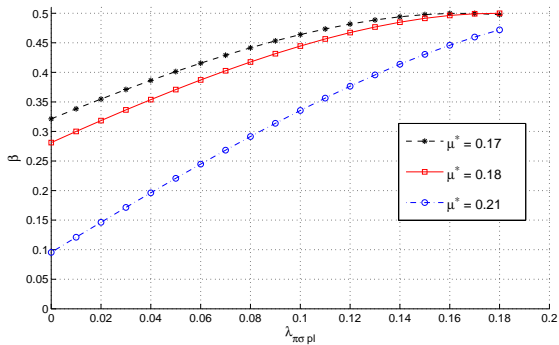}}}
\caption{Variation of  isotope exponent ($\beta$) with interband electron-plasmon  interaction ($\lambda_{\pi\sigma pl}$) for different values of $\mu^{\ast}$}
\label{isotopeVsPlasmon}
\end{figure}
\begin{figure}[!]
{\scalebox{1.0}
{\includegraphics[width=10cm]{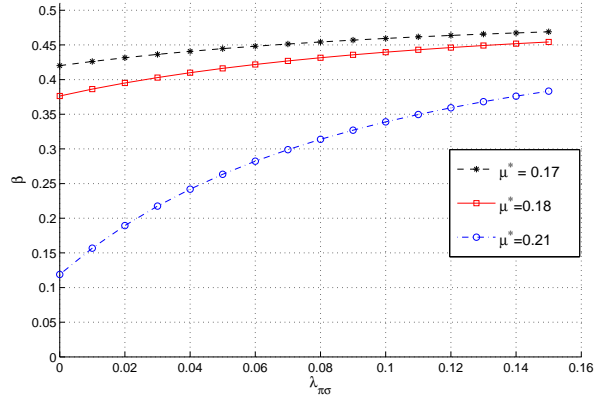}}}
\caption{Variation of  isotope exponent ($\beta$) with interband electron-phonon  interaction ($\lambda_{\pi\sigma}$) for different values of $\mu^{\ast}$}
\label{isotopeVsPhonon}
\end{figure}
\end{sloppypar}

\begin{sloppypar}
Figure (\ref{isotopeVsPlasmon}) shows the variation of isotope effect exponent, $\beta$ with the interband electron-plasmon interaction, $\lambda_{\pi\sigma pl}$ for a set of parameter . We can see that $\beta$ increase with $\lambda_{\pi\sigma pl}$ but decrease for increasing Coulomb pseudopotential. The variation of $\beta$ with $\lambda_{\pi\sigma}$ for various $\mu^{\ast}$ is shown in figure \ref{isotopeVsPhonon}. Both the interband contributions of phonon and collective excitations increase with the isotope exponent and tend to saturate at BCS predicted value of 0.5 for conventional superconductors. 
\end{sloppypar}

\section{Conclusion}

We have formulated a two-band model within the Bogoliubov-Valatin \cite{Bogoliubov,Valatin} formalism and by incorporating the effect of the collective excitation in the system, we  estimated the transition temperature, $T_c \simeq 38.80$ K and the isotope effect exponent, $\beta \simeq 0.43$. Our analysis shows that the inclusion of electron-phonon and electron-plasmon enhances the transition temperature of a superconducting MgB$_2$. Although, the model fails to account for the experimental observed value of $\beta$, it gives evidence that pairing mechanism in MgB$_2$ is not purely phonon and the effect of interband contributions are not negligible. Thus, more work is needed both theoretical and experimentally to understand the effect of this non-phonon mechanism. This is because only isotope effect cannot be used to assign the form of pairing mechanism in a superconducting material and the reduced isotope effect exponent of superconducting MgB$_2$ is still unclear \cite{Brotto2008}. 

\acknowledgements
We thank O. Umeh for reading the manuscript.

{}

\end{document}